# Joint Doppler frequency shift compensation and data detection method using 2-D unitary ESPRIT algorithm for SIMO-OFDM railway communication systems

Huiyue Yi, *Member, IEEE*

**Abstract—In this paper, we present a joint Doppler frequency shift compensation and data detection method using 2-D unitary ESPRIT algorithm for SIMO-OFDM railway communication systems over fast time-varying sparse multipath channels. By creating the spatio-temporal array data matrix utilizing the ISI-free part of the CP (cyclic prefix), we first propose a novel algorithm for obtaining auto-paired joint DOA and Doppler frequency shift estimates of all paths via 2-D unitary ESPRIT algorithm. Thereafter, based on the obtained estimates, a joint Doppler frequency shift compensation and data detection method is developed. This method consists of three parts: (a) the received signal is spatially filtered to get the signal corresponding to each path, and the signal corresponding to each path is compensated for the Doppler frequency shift in time domain, (b) the Doppler frequency shift-compensated signals of all paths are summed together, and (c) the desired information is detected by performing FFT on the summed signal after excluding the CP. Moreover, we prove that the channel matrix becomes time-invariant after Doppler frequency shift compensation and the ICI is effectively avoided. Finally, simulation results are presented to**

This work was supported in part by the National key project under Grant 2011ZX03001-007-03 and 2012ZX03003006-003, in part by the Science and Technology Breakthrough Plan of Shanghai Science and Technology Committee under Grant 10DJ1400302.
Huiyue Yi is with the Shanghai Research Center for Wireless Communications (WiCO) and Key Laboratory of Wireless Sensor Networks & Communication, Shanghai Institute of Microsystem and information technology, Chinese Academy of Science, 6th floor, Lane. 280-1, Linhong Road, Changning District, Shanghai 200335, P. R. China (e-mail: yi_huiyue@yahoo.com.cn; huiyue.yi@mail.sim.ac.cn)

.

**demonstrate the performance of the proposed method and compare it with the conventional method.**

*Index Terms*—**Unitary ESPRIT; direction of arrival; array signal processing; beamforming; Doppler frequency shift; spatial filtering**

I. INTRODUCTION

Along with the developments of the high-speed railway in many countries, the conventional "Global System for Mobile Communications Railways" (GSM-R) [1]-[2] can't meet the requirement for accessing high bandwidth for broadband multimedia services with sufficient level of QoS from increasing number of passengers on high-speed trains. In order to meet this requirement, a Wi-Fi based communication system was introduced for the ground-train connection [3]. Though Wi-Fi based system is attractive in terms of the cost efficiency, it is not easy to maintain the communication quality because the coverage of each wireless access point (AP) is small and the mobile trains experience frequent Layer 2 and Layer 3 handovers. With the increasing of train speed and demanding of data rate, it is getting more and more important to provide a better link quality in high-mobility communication systems. Broadband access on the train is provided through the train access terminal (TAT) [4]-[5], where data from all passengers in the train are first collected by the TAT, and then though the antenna on the top of the train delivered to the ground base station (BS). Recently, the "Long Term Evolution for Railway" (LTE-R) becomes a promising solution to provide high data rate in high-speed railway communication [6]. The OFDM and MIMO are two core techniques of the LTE systems.

In fading channels with very high mobility, the time-varying wireless channels with large Doppler spread may change at each OFDM symbol time [7]-[12]. This time variation of the channel over an OFDM symbol period results in a loss of subcarrier orthogonality and thus will lead to inter-carrier interference (ICI) due to power leakage among OFDM subcarriers [13]-[14].

As shown in [8], the time-varying multipath channel produces a time-varying complex multiplier at each subcarrier. Consequently, tracking time-varying channels with large Doppler spreads is a critical task to mitigate the ICI. There currently exist algorithms for channel estimation in mobile MIMO-OFDM systems with large Doppler shifts [10]-[12]. However, these algorithms all require the use of extensive preambles or training sequences in the time–frequency domain, which will decrease the spectrum efficiency. As is well known, array signal processing techniques are widely employed to acquire spatial direction-of-arrivals (DOA) of moving targets [15]-[16] and applied in wireless communications [17]-[18]. In [19], a joint DOA-frequency offset estimation and data detection method was introduced for the uplink MIMO-OFDM networks with SDMA techniques. In [20], the authors introduced an algorithm for joint estimation of the parameters including the frequency offsets, delays, and the angle selectivity in the uplink of multiuser MIMO-OFDM interference network. However, the method in [19] does not consider the time delay of different paths, while the algorithm in [20] considers that the effect of the Doppler frequency spreading on the frequency offsets of different paths of the same user are negligible. Therefore, the methods in [19]-[20] can not apply to the high-speed railway channel with large delay spread and rapid Doppler transition which can't be neglected.

Recently, lots of research works [21]-[25] have shown the channels in broadband wireless communication systems can often be modeled as a sparse channel, where the delay spread could be very large, but the number of dominant paths is normally very small. In a viaduct for the high-speed railway environment where the trackside base station (BS) is about 10-30 meters away from the railway track [26], the scatters are limited in this environment, and thus the specular LOS components are much stronger and other multipath echoes are encountered less frequently between the mobile train and the trackside BS, which provides a benign environment for the broadband mobile radio provision. In [27], the railway communication channel model is modeled as a sparse time-variant multipath channel. In this paper, we model the railway communication channel as a sparse fast time-varying multipath channels [7], and focus on investigating Doppler frequency shift compensation and data detection method for SIMO-OFDM railway communication systems. For SIMO-OFDM systems, some solutions to combat ICI have been proposed [28]-[29], which suppress Doppler-induced ICI via receive beamforming or adequate combining of the antenna signals. However, these approaches exploit only the statistical properties of the ICI and therefore can only partially mitigate the ICI effects. In [30], the authors introduced an ICI compensation technique for SIMO systems to combat the instantaneous Doppler-induced ICI distortions in the frequency domain. Since the FFT introduces the ICI because of Doppler frequency shift, there will be an irreducible error floor for data detection via frequency-domain Doppler frequency shift suppression method in [30] when the normalized Doppler frequency shift is relatively high. Consequently, the Doppler frequency shift should be compensated in the time domain before the FFT. As a result, in this paper we will focus on investigating Doppler frequency shift compensation methods in time domain before the FFT.

Utilizing array signal processing techniques, in this paper we propose an efficient joint Doppler frequency shift compensation and data detection method using 2-D unitary ESPRIT algorithm [31]-[32] for SIMO-OFDM railway communication systems over fast time-varying sparse multipath channels. We first propose a novel efficient algorithm for obtaining joint DOA and Doppler frequency shift estimates of each path for sparse multipath channels by utilizing 2-D unitary ESPRIT algorithm [31]-[32]. In the proposed algorithm, we begin with creating spatio-temporal array data matrix in a way as in [33] by utilizing the ISI-free part of the CP (cyclic prefix), and then obtain the auto-paired joint DOA and Doppler frequency shift estimates of each path for sparse multipath channels via 2-D unitary ESPRIT algorithm. Thereafter, based on the auto-paired joint DOA and Doppler frequency shift estimates of each path, a joint Doppler frequency compensation and data detection method is developed. This method mainly consists of the following two parts: (a) the received signal is spatially filtered to get the signal corresponding to each path, and the signal corresponding to each path is compensated for the Doppler frequency shift, and (b) the Doppler frequency shift-compensated signals of all paths are summed together in time domain, and then the desired information is detected by performing FFT on the summed signal after excluding the CP. Moreover, we prove that the channel matrix becomes time-invariant after performing Doppler frequency shift compensation and the ICI is effectively avoided. Since the Doppler frequency shift is compensated in the time domain before the FFT, the Doppler frequency shift will not affect the performance of the proposed joint Doppler frequency shift compensation and data detection method, and therefore the proposed method outperforms the data detection method [30] via frequency-domain Doppler frequency shift suppression in frequency domain and the conventional methods [8], [11]-[12]. Finally, simulation results are presented to demonstrate the performance of the proposed joint Doppler frequency shift compensation and data detection method and compare it with the conventional method.

The organization of this paper is as follows. In Section Ⅱ, we first introduce the system model of the SIMO-OFDM railway communication system in fast time-varying sparse multipath channels. Then, we introduce the conventional method, and describe its shortcomings. Finally, we present a novel joint Doppler frequency shift compensation and data detection method using 2-D unitary ESPRIT algorithm for SIMO-OFDM railway communication systems. Simulation results and conclusions are given in Sections Ⅲ and Ⅳ, respectively.

## II. JOINT DOPPLER FREQUENCY SHIFT COMPENSATION AND DATA DETECTION METHOD USING 2-D UNITARY ESPRIT ALGORITHM FOR SIMO-OFDM RAILWAY COMMUNICATION SYSTEMS

In this section, we first introduce the system model for SIMO-OFDM railway communication systems. Then, we briefly introduce the conventional data detection method for SIMO-OFDM communication systems, and describe its disadvantages. Finally, we propose a novel joint Doppler frequency shift compensation and data detection method using 2-D unitary ESPRIT algorithm for SIMO-OFDM railway communication systems, and describe its advantages over the conventional methods.

A. *System model for SIMO-OFDM railway communication systems*

Fig. 1 shows the system model for the SIMO-OFDM railway communication systems, which consists of an antenna array of $M$ receive antennas at the trackside BS and one transmit antenna mounted on the top of a moving train. In Fig. 1, $D$ is the distance between the trackside BS and the railway, and $\theta_1$ corresponds to the DOA of the LoS path with $\mathrm{tg}(\theta_1) = D/r$.

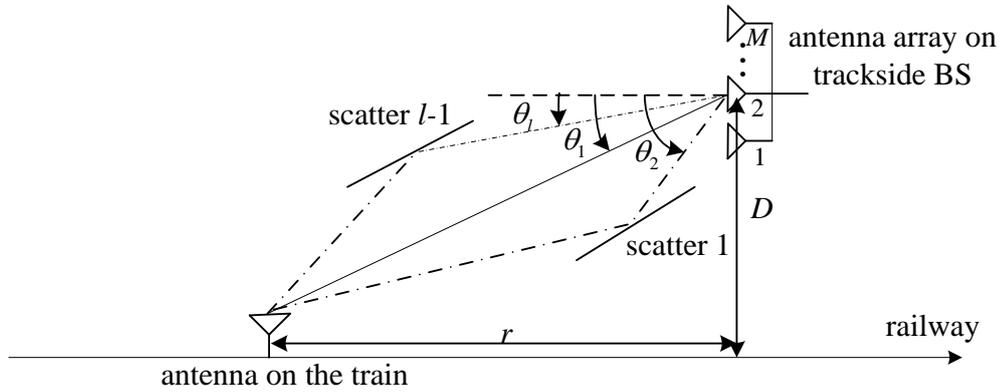

Fig. 1. System model for SIMO-OFDM railway communication systems

In OFDM multicarrier systems, the information symbols to be transmitted are first grouped into blocks of $N_c$ data symbols at the transmitter on the train, and the $i$ th block is represented by the vector $\mathbf{d}_i = [d_i(0), d_i(1), \cdots, d_i(N_c - 1)]^T$. The data symbols $d_i(k)$ from a certain modulation constellation in a finite symbol set can be assumed to be equi-probable and statistically independent of different subcarriers and blocks. Thus, $d_i(k)$ can be approximated as random variables with zero mean and correlation $E[d_i(j)d^*_{i'}(j')] = \sigma_d^2 \delta(j-j')\delta(i-i')$, where $\sigma_d^2$ is the power of data symbols and * is the conjugate operator. Each block is called a frequency-domain

OFDM symbol. In order to eliminate interference between parallel data streams, an $N_c$ point inverse fast Fourier transform (IFFT) is applied to this block. Then, a cyclic prefix (CP) of length $N_g$ as a copy of the last part of the IFFT output is inserted at the beginning of each symbol as the guard interval (GI) to avoid inter-symbol interference (ISI), and its length $N_g$ is assumed to be not shorter than the channel length. Thus, in discrete time, the baseband multicarrier signal transmitted by the antenna on the train can be written as follows:

$$s_i(n) = \sum_{k=0}^{N_c-1} d_i(k) \cdot e^{j2\pi nk/N_c}, \quad -N_g \leq k \leq N_c - 1 \tag{1}$$

When the train is moving at a speed of $v$, the Doppler frequency shift of the received signal will be given by

$$f_d = \frac{f_c v}{c}\cos(\theta) = \frac{v}{\lambda_c}\cos(\theta) \tag{2}$$

where $f_c$ is the carrier frequency, $\lambda_c$ is the carrier wavelength of the received signal, and $\theta$ denotes the angle between the direction of the moving train and the direction of the trackside BS. Then, the Doppler frequency shifted OFDM symbol is given by [9]

$$s'_i(n) = \sum_{k=0}^{N_c-1} d_i(k) e^{[j2\pi(nk/N_c)+j2\pi n f_d T_s]} \tag{3}$$

where $T_s$ is the duration of the original data symbol. The OFDM is sensitive to Doppler frequency shift and the system performance deteriorates with an increase in the Doppler frequency shift.

As illustrated in [26]-[27], the railway communication channel can be modeled as a fast time-varying sparse multipath channel between the trackside BS and the high-speed train. We assume the channel varies during the duration of an OFDM symbol due to the Doppler frequency shift. Let $Q$ be the number of resolvable paths in this multipath environment. Each path is parameterized by DOA $\theta_l$, time delay $\tau_l$ (measured in unit of the symbol periods $T_s$), the Doppler frequency shift $f_{d,l}$, and complex path attenuation factor $\beta_i(l)$, which is assumed stationary within an OFDM symbol period but varying between OFDM symbols. In addition, the maximum channel delay is denoted as $\tau_{\max} = \max\limits_{1 \le l \le Q} \tau_l$. We assume $\theta_1$ corresponds to the DOA of the LoS path with $\text{tg}(\theta_1) = D/r$, as can be seen from Fig. 1. Under the above assumptions, the channel can be modeled as [7]

$$\mathbf{H}_i(t) = \sum_{l=1}^{Q} \mathbf{a}(\theta_l)\beta_i(l)e^{(j2\pi f_{d,l}t)}\delta(t - \tau_l) \tag{4}$$

where $\mathbf{a}(\theta_l)$ is the $M \times 1$ array response to a path from the direction $\theta_l$, which for the uniform linear array (ULA) has the form

$$\mathbf{a}(\theta_l) = [1, e^{-j2\pi \frac{d}{\lambda_c}\sin(\theta_l)}, \cdots, e^{-j2\pi(M-1)\frac{d}{\lambda_c}\sin(\theta_l)}]^T \tag{5}$$

where $d$ is the inter-element spacing of the antenna array. During the transmission, the transmitted signal $s_i(n)$ passes through the multipath channel with the impulse response $\mathbf{H}_i(t)$, and gets corrupted by a spatially uncorrelated additive white Gaussian noise. In the $i$th OFDM block, the discrete-time received signal vector at the antenna array of the trackside BS can be expressed as the convolution of the transmitted signal $s_i(n)$ with the channel matrix $\mathbf{H}_i(t)$:

$$\begin{aligned}\mathbf{y}_i(n) &= \sum_{l=1}^{Q}\beta_i(l)\mathbf{a}(\theta_l)s'_i(n-\tau_l) + \mathbf{v}(n) \\ &= \sum_{l=1}^{Q}\{\beta_i(l)\mathbf{a}(\theta_l)e^{j2\pi(n-\tau_l)f_{d,l}T_s}\sum_{k=0}^{N_c-1}[d_i(k)e^{\frac{j2\pi(n-\tau_l)k}{N_c}}]\} + \mathbf{v}(n) \\ &= \sum_{k=0}^{N_c-1}d_i(k)e^{\frac{j2\pi nk}{N_c}}\sum_{l=1}^{Q}[\beta_i(l)\mathbf{a}(\theta_l)e^{j2\pi(n-\tau_l)\frac{f_{d,l}}{N_c \Delta f}}e^{\frac{-j2\pi\tau_l k}{N_c}}] + \mathbf{v}(n)\end{aligned} \tag{6}$$

where $\mathbf{y}_i(n)$ is the $n$th sample of the antenna array output vector, and $\mathbf{v}(n)$ is a $M \times 1$ AWGN vector with covariance matrix $\sigma^2 \mathbf{I}_M$ ($\mathbf{I}_M$ is the $M \times M$ identity matrix). Let us define

$$\mathbf{H}_{i,k}(n) = \sum_{l=1}^{Q} [\beta_i(l)\mathbf{a}(\theta_l) e^{j2\pi(n-\tau_l)\frac{f_{d,l}}{N_c \Delta f}} e^{-j2\pi\tau_l k/N_c}] \tag{7}$$

At the receiver, in the range $0 \leq n \leq N_c - 1$ of the $i$th OFDM symbol, the received signal is not corrupted by the previous OFDM symbols due to the presence of the CP. Then, $\mathbf{y}_i(n)$ can be rewritten as

$$\mathbf{y}_i(n) = \sum_{k=0}^{N_c-1} d_i(k)\mathbf{H}_{i,k}(n)e^{j2\pi nk/N_c} + \mathbf{v}(n), \quad 0 \leq n \leq N_c - 1 \tag{8}$$

Our aim is to efficiently detect the desired information symbol $\mathbf{d}_i = [d_i(0), d_i(1), \cdots, d_i(N_c - 1)]^T$ form the received signal $\mathbf{y}_i(n)$.

*B. Conventional data detection approach*

In the conventional receiver [8], the demodulation is performed using the FFT after excluding the guard interval CP. Then, the output for the $m$th subcarrier of the $i$th block can be expressed as

$$\begin{aligned}\mathbf{Y}_{i,m} &= \frac{1}{N_c} \sum_{n=0}^{N_c-1} \mathbf{y}_i(n)e^{-j2\pi nm/N_c} \\ &= \frac{1}{N_c} \sum_{n=0}^{N_c-1} [\sum_{k=0}^{N_c-1} d_i(k)\mathbf{H}_{i,k}(n)e^{\frac{j2\pi nk}{N_c}}]e^{-\frac{j2\pi nm}{N_c}} + \mathbf{V}_m \\ &= \frac{1}{N_c} \sum_{k=0}^{N_c-1} [d_i(k)\sum_{n=0}^{N_c-1} \mathbf{H}_{i,k}(n)e^{j2\pi(k-m)n/N_c}] + \mathbf{V}_m \end{aligned} \tag{9}$$

where $\mathbf{V}_m = \sum_{n=0}^{N_c-1} \mathbf{v}(n)e^{-j2\pi nk/N_c}$. From (9), we obtain the $k$th subcarrier of the $i$th block as follows:

$$\mathbf{Y}_{i,k} = \mathbf{H}_{i,k} d_i(k) + \mathbf{\eta}_k + \mathbf{V}_k \tag{10}$$

where $\mathbf{H}_{i,k} = (1/N_c)\sum_{n=0}^{N_c-1} \mathbf{H}_{i,k}(n)$, and $\mathbf{\eta}_k = (1/N_c)\sum_{m=0,m\neq k}^{N_c-1} [d_i(m)\sum_{n=0}^{N_c-1} \mathbf{H}_{i,k}(n)e^{j2\pi(k-m)n/N_c}]$. The $\mathbf{\eta}_k$ represents the ICI caused by the effect of the Doppler frequency shift. When the maximal normalized Doppler frequency is relatively small, the power of the ICI $\mathbf{\eta}_k$ can be neglected compared to the background noise power. In this case, the desired signal can be recovered as follows:

$$\hat{d}_i(k) = \mathbf{H}_{i,k}^H \mathbf{Y}_{i,k} / (\mathbf{H}_{i,k}^H \mathbf{H}_{i,k}) = d_i(k) + \mathbf{H}_{i,k}^H (\boldsymbol{\eta}_k + \mathbf{V}_k) / (\mathbf{H}_{i,k}^H \mathbf{H}_{i,k}) \tag{11}$$

Because the channel matrix $\mathbf{H}_{i,k}$ is time-varying, so it requires intensive pilot subcarriers to estimate $\mathbf{H}_{i,k}$, which will decrease the spectrum efficiency. Moreover, since the Doppler frequency shift destroys the orthogonality between subcarriers, the FFT introduces ICI $\boldsymbol{\eta}_k$. Consequently, there will be an irreducible error floor for the conventional method when the normalized Doppler frequency shift becomes relatively high.

*C. Joint Doppler frequency shift compensation and data detection method using 2-D unitary ESPRIT algorithm*

In this subsection, we present a novel joint Doppler frequency shift compensation and data detection method using array signal processing techniques for the SIMO-OFDM railway communication systems. The proposed method consists of two stages: (1) joint DOA and Doppler frequency shift estimation using 2-D unitary ESPRIT algorithm; (2) Doppler frequency shift compensation and data detection.

*1) Joint DOA and Doppler frequency shift estimation algorithm using 2-D unitary ESPRIT*

In this subsection, we propose a joint DOA and Doppler frequency shift algorithm by utilizing 2-D unitary ESPRIT algorithm [31]-[32]. In vector form, $\mathbf{y}_i(n)$ in (6) can be rewritten as

$$\begin{aligned}\mathbf{y}_i(n) &= \sum_{l=1}^{Q} \{\beta_i(l)\mathbf{a}(\theta_l) e^{j2\pi(n-\tau_l)\frac{f_{d,l}}{N_c \Delta f}} s_i(n-\tau_l)\} + \mathbf{v}_0(n) \\ &= \mathbf{ABd}_i(n) + \mathbf{v}_0(n)\end{aligned} \tag{12}$$

where $\mathbf{A} = [\mathbf{a}(\theta_1), \mathbf{a}(\theta_2), \cdots, \mathbf{a}(\theta_Q)] \in C^{M \times Q}$, $\mathbf{B} = \mathrm{diag}\{\beta_i(1)\,\beta_i(2)\cdots\beta_i(Q)\}$ is the $Q \times Q$ fading matrix, $\mathbf{d}_i(n) = \mathbf{g}_i(n) \odot \mathbf{s}_i(n)$ is the $Q \times 1$ temporal signature matrix, where $\odot$ denotes the

11Hadamard product, $\mathbf{g}_i(n) = [1, e^{j2\pi(n-\tau_1)\frac{f_{d,1}}{N_c\Delta f}}, \cdots, e^{j2\pi(n-\tau_Q)\frac{f_{d,Q}}{N_c\Delta f}}]^T$, and $\mathbf{s}_i(n) = [s_i(n-\tau_1)\ s_i(n-\tau_2)\cdots s_i(n-\tau_Q)]^T$. From (12), we can get

$$\begin{aligned}
\mathbf{y}_i(n+N_c) &= \sum_{l=1}^{Q}\{\beta_i(l)\mathbf{a}(\theta_l)e^{j2\pi\frac{f_{d,l}}{\Delta f}}e^{j2\pi(n-\tau_l)\frac{f_{d,l}}{N_c\Delta f}}s_i(n+N_c-\tau_l)\} + \mathbf{v}_0(n+N_c) \\
&= \mathbf{A}\mathbf{\Phi}\mathbf{B}\mathbf{g}_i(n)\odot\mathbf{s}_i(n+N_c) + \mathbf{v}_0(n+N_c)
\end{aligned} \quad (13)$$

where $\mathbf{\Phi} = \mathrm{diag}\{e^{j2\pi f_{d,1}/\Delta f}, e^{j2\pi f_{d,2}/\Delta f}, \cdots, e^{j2\pi f_{d,Q}/\Delta f}\}$. We partition the CP of length $N_g$ into two parts: ISI-contaminated part of length $\tau_{\max}$, and ISI-free part of length $P$, i.e., $N_g = \tau_{\max} + P$. Since $s_i(n) = s_i(N_c+n)$ ($n = -N_g, \cdots, -1$), then in the ISI-free part we have

$$\mathbf{s}_i(n) = \mathbf{s}_i(n+N_c), \qquad n = -P, \cdots, -1. \quad (14)$$

As a result, we obtain

$$\begin{aligned}\mathbf{y}_i(n+N_c) \\ = \mathbf{A}\mathbf{\Phi}\mathbf{B}\mathbf{d}_i(n) + \mathbf{v}_0(n+N_c)\end{aligned}, \quad n = -P, \cdots, -1 \quad (15)$$

In order to jointly estimate the DOA and Doppler frequency shift of each path via 2-D unitary ESPRIT [31]-[32], we create the following spatio-temporal array data matrix in a similar way as in [33]:

$$\begin{aligned}
\mathbf{Y}_i &= \begin{bmatrix} \mathbf{y}_i(-P) & \mathbf{y}_i(-P+1) & \cdots & \mathbf{y}_i(-1) \\ \mathbf{y}_i(N_c-P) & \mathbf{y}_i(N_c-P+1) & \cdots & \mathbf{y}_i(N_c-1) \end{bmatrix} \\
&= \begin{bmatrix} \mathbf{A} \\ \mathbf{A}\mathbf{\Phi} \end{bmatrix}\mathbf{B}\mathbf{D}_i + \mathbf{V} \\
&= \mathbf{\Omega}\mathbf{B}\mathbf{D}_i + \mathbf{V} \in C^{2M\times P}
\end{aligned} \quad (16)$$

where $\mathbf{\Omega} \in \mathbb{C}^{2M\times Q}$ is referred to as the spatio-temporal array manifold matrix, $\mathbf{V} \in \mathbb{C}^{2M\times P}$ is the noise matrix, and $\mathbf{D}_i = [\mathbf{d}_i(-P)\ \mathbf{d}_i(-P+1), \cdots, \mathbf{d}_i(-1)]$. The conditions on the parameters $M$, $P$ in (16) for guaranteeing the resolvability of signals of $Q$ paths are analyzed thoroughly in [33], and thus is omitted here. Based on the data matrix $\mathbf{Y}_i$ in (16), the auto-paired $(\hat{\theta}_l, \hat{f}_{d,l})$ estimates $(l = 1, 2, \cdots, Q)$ can be efficiently obtained using the 2-D unitary ESPRIT algorithm in [31]-[32].

*2) Doppler frequency shift compensation and data detection method*

Based on auto-paired $(\hat{\theta}_l, \hat{f}_{d,l})$ estimates, we propose a novel Doppler frequency shift compensation and data detection method in this subsection. For simplifying the exposition of the proposed method, we assume that $(\hat{\theta}_l, \hat{f}_{d,l})$ of all paths are perfectly estimated, i.e., $\hat{\theta}_l = \theta_l$, and $\hat{f}_{d,l} = f_{d,l}$ ($l = 1, 2, \cdots, Q$). The effect of estimation errors of $(\theta_l, f_{d,l})$ on the performance of the proposed method will be evaluated through simulations. In order to filter out the signal corresponding to the $l$ th path, we construct $Q$ spatially-filtering matrices as follows [19]:

$$\mathbf{F}_l = \mathbf{I} - \mathbf{T}_l \mathbf{T}_l^\dagger, \quad l = 1, 2, \cdots, Q \tag{17}$$

where $\mathbf{T}_l = [\mathbf{a}(\hat{\theta}_1), \cdots, \mathbf{a}(\hat{\theta}_{l-1}), \mathbf{a}(\hat{\theta}_{l+1}), \cdots, \mathbf{a}(\hat{\theta}_Q)]$, and $(\cdot)^\dagger$ is the Moore-Penrose pseudo inverse. Then, the spatially-filtered signals for the $l$ th path is given by

$$\begin{aligned}\mathbf{z}_l(n) &= \mathbf{F}_l \mathbf{y}_i(n) \\ &= \beta_i(l)\mathbf{a}(\theta_l)e^{j2\pi(n-\tau_l)\frac{f_{d,l}}{N_c\Delta f}}s_i(n-\tau_l) + \mathbf{F}_l \mathbf{v}_0(n)\end{aligned}, \quad l = 1, 2, \cdots, Q \tag{18}$$

The spatially filtered signal $\mathbf{z}_l(n)$ only consists of the desired signal form the DOA $\hat{\theta}_l$ of the $l$th path. After the spatially filtered signal $\mathbf{z}_l(n)$ is multiplied with $\exp[-j2\pi n\hat{f}_{d,l}/N_c\Delta f]$ to compensate for the Doppler frequency shift $f_{d,l}$, we obtain

$$\begin{aligned}\mathbf{x}_l(n) &= e^{-j2\pi n\hat{f}_{d,l}/N_c\Delta f}\mathbf{z}_l(n) \\ &= \beta_i(l)\mathbf{a}(\theta_l)e^{-j2\pi\tau_l\frac{f_{d,l}}{N_c\Delta f}}s_i(n-\tau_l) + e^{-j2\pi n\hat{f}_{d,l}/N_c\Delta f}\mathbf{F}_l\mathbf{v}_0(n) \\ &= \beta_i(l)\mathbf{a}(\theta_l)e^{-j2\pi\tau_l\frac{f_{d,l}}{N_c\Delta f}}s_i(n-\tau_l) + \mathbf{v}'_l(n)\end{aligned} \tag{19}$$

Then, we compute the sum of the Doppler-frequency compensated signal $\mathbf{x}_l(n)$ of all paths in the time domain as follows:

$$\begin{aligned}
\mathbf{x}_c(n) &= \sum_{l=1}^{Q} \mathbf{x}_l(n) \\
&= \sum_{l=1}^{Q} \beta_i(l)\mathbf{a}(\theta_l) e^{-j2\pi\tau_l \frac{f_{d,l}}{N_c \Delta f}} \sum_{k=0}^{N_c-1} [d_i(k) e^{j2\pi(n-\tau_l)k/N_c}] + \sum_{l=1}^{Q} \mathbf{v}'_l(n) \\
&= \sum_{k=0}^{N_c-1} d_i(k) \{ \sum_{l=1}^{Q} \beta_i(l)\mathbf{a}(\theta_l) e^{-j2\pi\tau_l(\frac{f_{d,l}}{N_c \Delta f}+\frac{k}{N_c})} \} e^{j2\pi nk/N_c} + \mathbf{v}_c(n) \\
&= \sum_{k=0}^{N_c-1} d_i(k) \bar{\mathbf{H}}_{i,k} e^{j2\pi nk/N_c} + \mathbf{v}_c(n)
\end{aligned} \quad (20)$$

where $\bar{\mathbf{H}}_{i,k} = \sum_{l=1}^{Q} \beta_i(l)\mathbf{a}(\theta_l) e^{-j2\pi\tau_l(\frac{f_{d,l}}{N_c \Delta f}+\frac{k}{N_c})}$. Compared with time-varying $\mathbf{H}_{i,k}(n)$ in (9) in the conventional method, $\bar{\mathbf{H}}_{i,k}$ in (20) becomes time-invariant in the duration of an OFDM symbol. Thus, the required number of pilot subcarriers for estimating $\bar{\mathbf{H}}_{i,k}$ will be less than that for estimating $\mathbf{H}_{i,k}$ in (10). The demodulation is performed on $\mathbf{x}_c(n)$ in (20) using the FFT after excluding the CP. Then, the output for the $m$th subcarrier of the $i$th block can be expressed as

$$\mathbf{z}_i(m) = \frac{1}{N_c} \sum_{n=0}^{N_c-1} \sum_{k=0}^{N_c-1} d_i(k) \bar{\mathbf{H}}_{i,k} e^{j2\pi n(k-m)/N_c} + \mathbf{v}'_c(m) \quad (21)$$

From (21), we obtain the output for $k$th subcarrier of the $i$th block as follows:

$$\mathbf{z}_i(k) = \bar{\mathbf{H}}_{i,k} d_i(k) + \mathbf{v}'_c(k) \quad (22)$$

Comparing with the conventional data detect method in (11) where there are ICI term $\boldsymbol{\eta}_k$ caused by the effect of the Doppler frequency shift, there is no ICI term in (22) because the Doppler frequency shifts have been effectively compensated in the time domain before the FFT. As a result, the Doppler frequency shift will not affect the performance of the proposed method, which will be demonstrated via simulations in Section Ⅲ. Utilizing maximum ratio combining (MRC), the frequency-domain symbol estimates $\hat{d}_i(k)$ can be obtained as:

$$\hat{d}_i(k) = \bar{\mathbf{H}}_{i,k}^{H} \mathbf{z}_i(k) / [\bar{\mathbf{H}}_{i,k}^{H} \bar{\mathbf{H}}_{i,k}] = d_i(k) + \bar{\mathbf{H}}_{i,k}^{H} \mathbf{v}'_c(k) / (\bar{\mathbf{H}}_{i,k}^{H} \bar{\mathbf{H}}_{i,k}) \quad (23)$$

Finally, we can retrieve the information bits from $\hat{d}_i(k)$.

## III. SIMULATIONS AND DISCUSSIONS

In this section, we demonstrate through computer simulations the performance of the proposed joint Doppler frequency shift compensation and data detection method using 2-D unitary ESPRIT for the SIMO-OFDM railway communication systems and compare it with the conventional method. In simulations, the number of receive antennas at the trackside BS is set to $M = 5$, and the receive antennas take the form of a ULA with an inter-element spacing of $\lambda_c/2$. The number of subcarriers $N_c$ is 512, the intercarrier spacing is 15kHz, and the maximum channel time delay is $\tau_{max} = 28$. Information symbols $d_i(n)$ ($i = 1, 2, \cdots, 512$) are independent and identically distributed (i.i.d.) QPSK symbols. For the multipath channel, we assume the number of path is $Q = 3$. The DOAs of three paths are $(\theta_1, \theta_2, \theta_3) = (1°, 35°, 60°)$, and the time delays are $(\tau_1, \tau_2, \tau_3) = (0, 2, 6)T_s$. The path attenuation factor for the three path are, respectively, $\beta_i(1) = e^{-j\phi_1}$, $\beta_i(2) = 0.6e^{-j\phi_2}$, and $\beta_i(3) = 0.36e^{-j\phi_3}$, where $\phi_l$ ($l = 1, 2, 3$) are random initial phases. The velocity of the train is $v = 360$km/h (i.e., 100m/s). In order to illustrate the effect of Doppler frequency shifts on the performance of the proposed method, the carrier frequency $f_c$ is set as 3GHz, 6GHz and 9GHz to produce various maximum Doppler frequency shifts. The maximum Doppler frequency shifts corresponding to 3GHz, 6GHz and 9GHz are, respectively, 1000Hz, 2000Hz, and 3000Hz.

As the performance of the proposed Doppler frequency shift compensation and data detection method depends on the accuracy of the auto-paired DOA and Doppler frequency shift estimates, in the following we firstly examine the performance of the proposed joint DOA and Doppler frequency shift estimation algorithm via 2-D unitary ESPRIT. Then, we provide simulation results to illustrate the performance of the proposed Doppler frequency shift compensation and data detection method.

*A. Performance of the proposed joint DOA and Doppler frequency shift estimation algorithm via 2-D unitary ESPRIT*

In this subsection, we provide simulation results to illustrate the performance of the proposed joint DOA and Doppler frequency shift estimation algorithm via 2-D unitary ESPRIT. In simulations, we set the carrier frequency $f_c = 9\text{GHz}$, and the Doppler frequency shifts $(f_{d,1}, f_{d,2}, f_{d,3}) = (3000, 2500, 1500)\text{Hz}$. Fig. 2 shows the mean values of the estimates for the DOA $\theta_1$ and Doppler frequency shifts $f_{d,1}$ versus $E_b/N_0$ for various length $P$ when $f_c = 9\text{GHz}$, and the dashed line with a mark * denotes the true values of $\theta_1$ and $f_{d,1}$. Fig. 3 shows the root mean square error (RMSE) of the estimates for the DOA $\theta_1$ and Doppler frequency shifts $f_{d,1}$ versus $E_b/N_0$ for various length $P$ when $f_c = 9\text{GHz}$. The corresponding results for $(\theta_2, f_{d,2})$ and $(\theta_3, f_{d,3})$ are similar and therefore not plotted in Fig. 2 and Fig. 3 for clarity. At each $E_b/N_0$, 2000 Monte Carlo independent trials were performed to obtain the statistical results.

As can be seen for Fig. 2, the mean values for the estimates of $\theta_1$ and $f_{d,1}$ tend to approach the true values as the SNR increases, and converge to the true values for $P = 25$ when $E_b/N_0 > 26\text{dB}$. It can be seen from the Fig. 3 that the RMSEs for the estimates of $\theta_1$ and $f_{d,1}$ tends to decrease as the SNR increases, and converge to zero for $P = 25$ when $E_b/N_0 > 26\text{dB}$. Moreover, the performance of the proposed joint DOA and Doppler frequency shift estimation algorithm improves as the length $P$ increases, and its performance improves about 4dB when $P$ increases from 25 to 100.

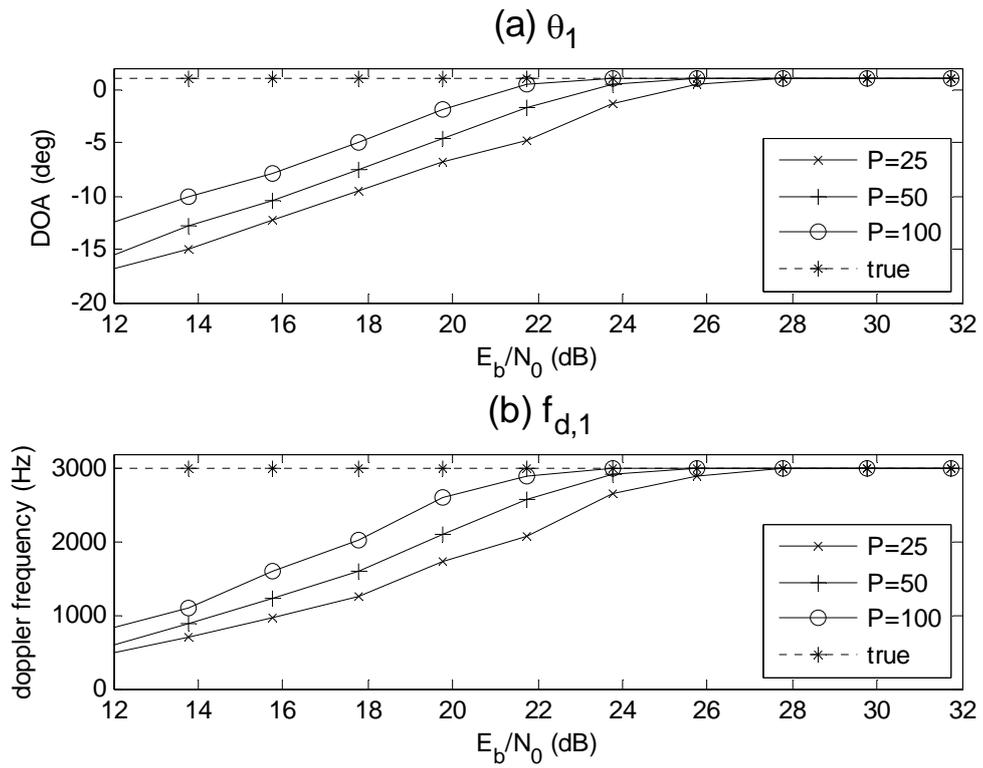

Fig. 2 Mean values of the estimates for the DOA $\theta_1$ and Doppler frequency shifts $f_{d,1}$ versus $E_b/N_0$ for various length $P$ when $f_c = 9$GHz

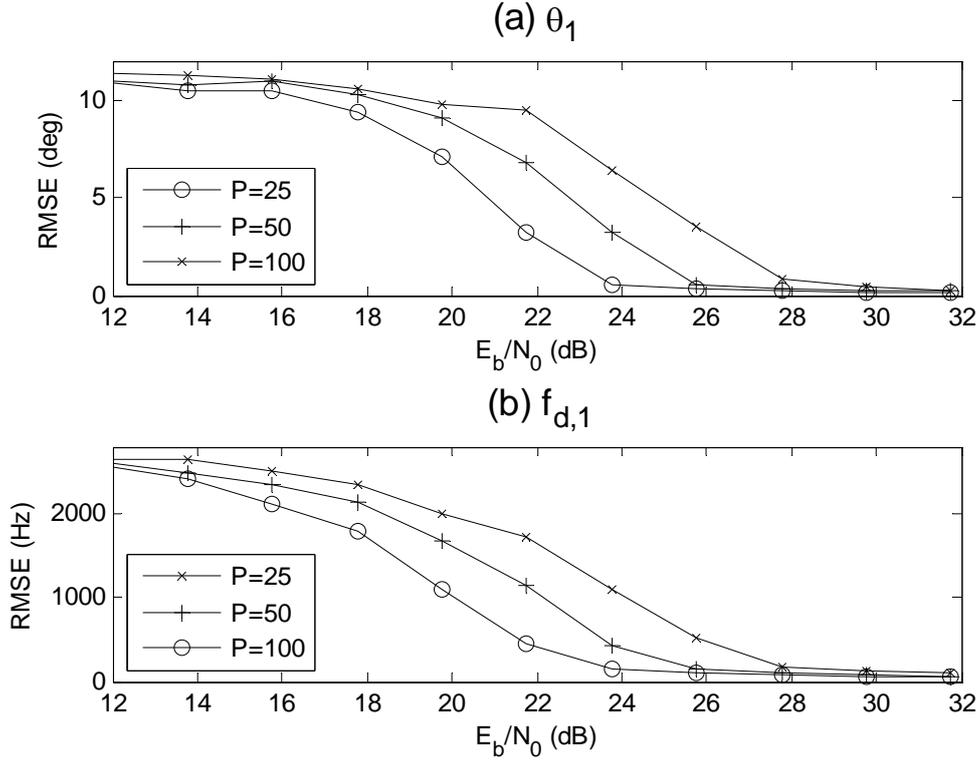

Fig. 3 Root mean square error (RMSE) of the estimates for the DOA $\theta_1$ and Doppler frequency shifts $f_{d,1}$ versus $E_b/N_0$ for various length $P$ when $f_c = 9$GHz

*B. Performance of the proposed Doppler frequency shift compensation and data detection method*

In this subsection, simulation results are presented to illustrate the performance of the proposed Doppler frequency shift compensation and data detection method and compare it with the conventional method. In order to focus on the effect of the joint DOA and Doppler frequency shift estimation algorithm on the performance of the proposed method, we assume that the channel matrix $\mathbf{H}_{i,k}$ in the conventional method (11) and $\bar{\mathbf{H}}_{i,k}$ in the proposed method (23) are perfectly known. In addition, we set the Doppler frequency shift $(f_{d,1}, f_{d,2}, f_{d,3}) = (3000, 2500, 1500)$Hz when $f_c = 9\text{GHz}$, $(f_{d,1}, f_{d,2}, f_{d,3}) = (2000, 1667, 1000)\text{Hz}$ when $f_c = 6\text{GHz}$, and $(f_{d,1}, f_{d,2}, f_{d,3}) = (1000, 833, 500)\text{Hz}$ when $f_c = 3\text{GHz}$.

Fig. 4 shows the BER performance of the proposed Doppler frequency shift compensation and data detection method versus $E_b/N_0$ for various carrier frequency $f_c$ when $P = 100$. The corresponding results for the proposed method with perfect $(\theta_i, f_{d,i})$ and the conventional method are also shown for comparison. Note that as the carrier frequency $f_c$ increases from 3GHz to 9GHz, the normalized maximum Doppler frequency shift increases from 1000/15000 (0.067) to 3000/15000 (0.2). As can be seen from Fig. 4, when the carrier frequency $f_c$ increases from 3GHz to 9GHz, the BER performance of the conventional method deteriorates significantly, while the BER performance of the proposed method with estimated (or perfect) $(\theta_i, f_{d,i})$ remains unchanged. As expected, the proposed method with perfect $(\theta_i, f_{d,i})$ outperforms the conventional method irrespective of Doppler frequency shifts. This is because the Doppler frequency shift of each path is effectively compensated in the time domain before FFT in the proposed method, and thus the Doppler frequency shift will not destroys the orthogonality between the subcarriers. Moreover, due to the estimation errors of $(\theta_i, f_{d,i})$ estimates, the performance of proposed method with estimated $(\theta_i, f_{d,i})$ ($P = 100$) degrades about 2dB compared to that with perfect $(\theta_i, f_{d,i})$ when $E_b/N_0 < 24\text{dB}$ and is very close to that with perfect $(\theta_i, f_{d,i})$ when $E_b/N_0 \geqslant 24\text{dB}$. This is because the estimation errors of $(\theta_i, f_{d,i})$ is relatively large when $E_b/N_0 < 24\text{dB}$ and is very slight when $E_b/N_0 \geqslant 24\text{dB}$ in the case of $P = 100$, as shown in Fig. 2 and Fig. 3.

In order to further demonstrate the performance of the proposed Doppler frequency shift compensation and data detection method, we examine the effect of various length $P$ of ISI-free part of the CP on the performance of the proposed method. Fig. 5 shows the BER performance of the proposed Doppler frequency shift compensation and data detection method versus $E_b/N_0$ for various length $P$ when the carrier frequency $f_c = 9\text{GHz}$. The corresponding results for the proposed method with perfect $(\theta_i, f_{d,i})$ are also shown for comparison. As can be seen from Fig. 5, the performance of the proposed method improves as the length $P$ increases. This is because the estimation accuracy of $(\theta_i, f_{d,i})$ estimates improves as the length $P$ increases, as shown in Fig. 2 and Fig. 3.

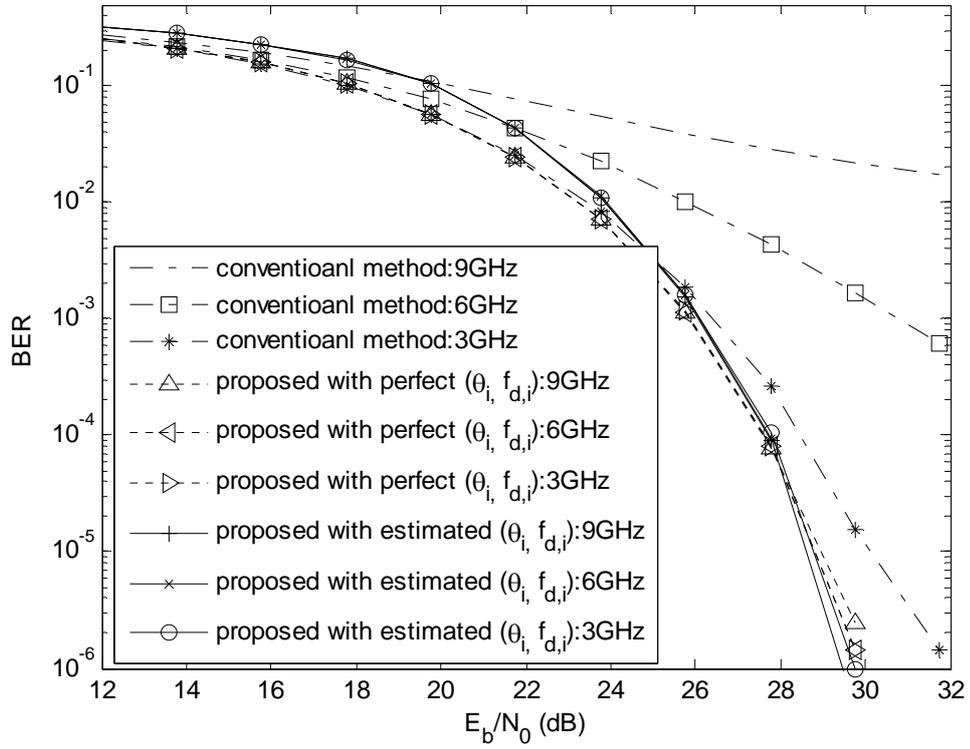

Fig. 4 BER performance of the proposed Doppler frequency shift compensation and data detection method versus $E_b/N_0$ for various carrier frequency $f_c$ when $P = 100$

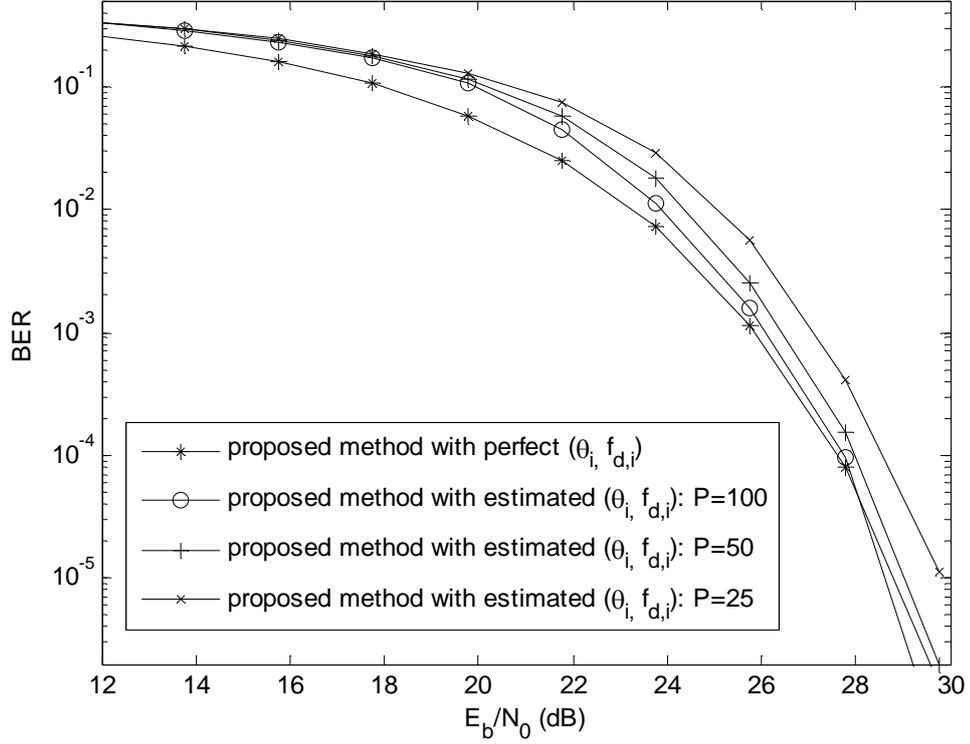

Fig. 5. BER performance of the proposed Doppler frequency shift compensation and data detection method versus $E_b/N_0$ for various length $P$ when the carrier frequency $f_c = 9\text{GHz}$

## IV. CONCLUSION

In this paper, we have presented a novel joint Doppler frequency shift compensation and data detection method using 2-D unitary ESPRIT algorithm for SIMO-OFDM railway communication systems over fast time-varying sparse multipath channels. We first proposed a novel algorithm for obtaining joint DOA and Doppler frequency shift estimates of all paths by using 2-D unitary ESPRIT algorithm. The proposed algorithm begins with creating spatio-temporal array data matrix by utilizing the ISI-free part of the CP (cyclic prefix), and then obtains the auto-paired joint DOA and Doppler frequency shift estimates of each path via 2-D unitary ESPRIT algorithm. Thereafter, based on the auto-paired joint DOA and Doppler frequency shift estimates, a joint Doppler frequency compensation and data detection method has been developed. This method mainly consists of two steps. Firstly, the received signal is spatially filtered to get the signal corresponding to each path, and the signal corresponding to each path is compensated for the Doppler frequency shift. Secondly, the Doppler frequency shift-compensated signals of all paths are summed together in the time domain, and the desired information is detected by performing FFT on the summed signal after excluding the CP. Moreover, we prove that the channel matrix becomes time-invariant after performing Doppler frequency shift compensation and the ICI is effectively avoided. Finally, it has been shown via simulations the effectiveness of the proposed joint DOA and Doppler frequency shift compensation algorithm as well as the proposed joint Doppler frequency shift compensation and data detection method. As expected, simulation results have shown that the performance of the proposed method is not affected by the Doppler frequency shift, and the proposed method exhibits much better BER performance than the conventional method when the normalized maximum Doppler frequency shift is relatively high.